\documentclass[preprint,showpacs,prc,aps,showkeys]{revtex4-1}
\usepackage{graphicx}
\usepackage{dcolumn}
\usepackage{bm}
\usepackage{hyperref}

\begin{document}

\preprint{APS/123-QED}

\title{Band crossing in Shears band of $^{108}$Cd }

\author{Santosh Roy}
\altaffiliation[Also at ]{Saha Institute of Nuclear Physics. 1/AF Bidhannager, Kolkata 700 064, India}
\affiliation{S. N. Bose National Centre for Basic Sciences. Block JD, Sector III, Saltlake City, Kolkata 700098, India}

\author{Pradip Datta}
\altaffiliation[Also at ]{Ananda Mohan College, 102/1 Raja Rammohan Roy Sarani, Kolkata 700 009, India}
\affiliation{ iThemba Labs,P.O. Box 722 Somerset West 7129, South Africa.}

\author{S. Pal}
\altaffiliation[Presently at]{ IRFU, CEA Saclay, 91191, Gif-sur-Yvette, France}
\author{ S. Chattopadhyay}
\author{S. Bhattacharya}
\author{A. Goswami}
\affiliation{Saha Institute of Nuclear Physics, 1/AF Bidhannager Kolkata, 700 064, India}

\author{H. C. Jain and P. K. Joshi}
\affiliation{Tata Institute of Fundamental Research, Homi Bhabha Road, Mumbai 
400 005, India}

\author{R. K. Bhowmik, R. Kumar, S. Muralithar, R. P. Singh, N. Madhavan}
\affiliation{Inter University Accelerator Center, Aruna Asaf Ali Marg, New Delhi 110 067, India}

\author{P. V. Madhusudhana Rao.}
\affiliation{Department of Physics, Andra University, Visakhapatnam 530 003, India}

\date{\today}

\begin{abstract}
The level lifetimes have been measured for a Shears band of $^{108}$Cd which exhibits bandcrossing.
The observed level energies and B(M1) rates have been successfully described by a semi-classical geometric model based on shear
mechanism. In this geometric model, the bandcrossing in Shears band has been described as the reopening of the angle between the blades of a shear.
 
\begin{description}

\item[PACS numbers] 21.10.Hw, 21.10.Re, 21.10.Tg, 21.60.Ev, 23.20.-g, 25.70.Gh, 27.60.+j

\end{description}
\end{abstract}

\keywords{Shears band, $^{108}$Cd, band crossing, lifetime measurement, semi-classical model}
\maketitle

In recent years a large number of Shears band have been identified in mass-100 region \cite{datta1, datta2, datta3,kelsall, clark1, clark2}. The shears structure in these nuclei originate due to proton holes in g$_{9/2}$ orbital and neutron particles in h$_{11/2}$/g$_{7/2}$/d$_{5/2}$ orbitals. The bands originating from shears mechanism exhibit sequences of magnetic-dipole
 (M1) transitions and thus, are often referred to as M1 bands. The observed Routhians and transition rates in these bands have been 
well described in the framework of Tilted Axis Cranking (TAC) \cite{fraun1, fraun2} which show that the total angular momentum is almost completely
 generated by the valance proton and neutron angular momenta. Thus, the different observed features of these Shears bands 
can also be described successfully by a semi-classical geometric model by Clark and Macchiavelli \cite{clark2, machia1} which
 involves the coupling of the two angular momentum vectors of protons and neutrons, namely \textbf{$\bm{j}_{\perp}$} and
 \textbf{$\bm{j}_{\parallel}$}. In this model, the observed band head spin is generated by the perpendicular coupling 
of \textbf{$\bm{j}_{\perp}$} and \textbf{$\bm{j}_{\parallel}$}. So, at the band head, the angle ($\theta$) between them
 (called shears angle) is 90$^\circ$. The higher spin states of the band originate due to gradual closing of
 these two vectors around their resultant which resembles the closing blades of a pair of shear and the
excitation energy along the band increases due to the potential energy associated with reorientation the two vectors.
 Thus, in this model the highest spin state for a given particle-hole configuration is obtained when the two vectors are fully aligned ($\theta=0$).

\indent The phenomenon of band crossing in Shears Band is quite similar to that found in the case of collective rotation, i.e at
 bandcrossing, the observed angular momentum is generated at a lower energy by a new higher 
quasi-particle configuration \cite{aks,priya,kelsall,cooper}. However, the two band crossings can be described in two different ways. In case of collective rotation, the bandcrossing 
is associated to the alignment of a pair of quasi-particles to the rotational axis whose angular momentum adds to the
 rotational angular momentum. Thus, if the energy needed to break the pair (coupled to zero angular momentum) is lower
 than that needed to generate the angular momentum through collective rotation, then the bandcrossing is observed. 
On the other hand, the bandcrossing in Shears band happens due to a new configuration which can reproduce the same 
angular momentum at a larger shear angle i.e at a lower energy. This happens because the magnitude of the vector(s) or the length(s)
 of the shear blade(s) increases due to participation of two more particles or holes in 
forming the new shear structure. The higher angular momentum states after the band crossing are then obtained 
by re-closing of the newly formed shear blades. Thus, the bandcrossing in Shears band can be related to the reopening of the shear angle.

\indent The validity of this geometric picture can be tested by the measurement of magnetic dipole transition B(M1) rates before
 and after the band crossing along a Shears band. This rate is proportional to the square of the perpendicular component of 
the magnetic moment and thus, decrease as the shears close \cite{clark2}. However, as the shear angle reopens after the band crossing, 
it is expected that the B(M1) rates will increase immediately after the band crossing followed by the characteristic drop
 which would indicate the closing of the new shears structure. This feature was observed in $^{196}$Pb \cite{aks} where the reopening of the shears angle was caused by the alignment of an i$_{13/2}$  neutron pair. Before the band crossing the B(M1) rates were found to drop from 2.4 to 0.7 ${\mu_N}^2$ while after the crossing the values increased to $\sim$ 9 ${\mu_N}^2$ and drop to 1.8 ${\mu_N}^2$ with increasing spin.

\indent In the mass-100 region, a band crossing in Shears band was first reported in $^{108}$Cd by Thorslund {\it et al.}\cite{thor1}. 
This negative parity band (labeled as Band 5 in \cite{thor1}) comprises of a sequence of ten M1 transitions namely, 121, 316, 522, 677, 466, 362, 482, 590, 706 and 798 keV. 
Thus, it is apparent that the smooth parabolic behavior of the Shears band is broken after 677 keV transition and a new sequence develops 
beyond 466 keV transition. The configuration before band crossing was suggested to be $\pi [{g_{9/2}}^{-2}]_{j=8} \otimes \nu [{h_{11/2}}^{1}(g_{7/2}/d_{5/2})^{1}]$ and the bandcrossing was assigned to $\nu (h_{11/2})^{2}$ alignment.
The lifetimes of the levels in this band were measured by Kelsall {\it et al.} \cite{kelsall}. In this work, $\pi[g_{9/2}^{-3}g_{7/2}]\otimes \nu[h_{11/2}(g_{7/2}/d_{5/2})]$ configuration was assigned to this band before the neutron alignment based on the observed bandhead energy and B(M1) rates. 
The measured B(M1) rates showed the characteristic fall as a function of angular momentum after the neutron alignment. 
However, this fall was not observed before the alignment. Thus, the phenomena of band crossing in Shears band could not be established in $^{108}$Cd. In the present work, we have re-measured the level lifetimes of this band by analyzing the observed
 lineshapes of all the transitions except 121 and 798 keV. 

\indent The high spin levels of $^{108}$Cd were populated through $^{100}$Mo ($^{13}$C, $5n\gamma$) $^{108}$Cd reaction using 65 MeV $^{13}$C beam from 15-UD Pelletron  at Inter University Accelerator Centre, New Delhi \cite{mehta} . 
The gamma rays were detected by an array of eight compton suppressed Clover detectors. The detectors were mounted on two opposite rings at nominal angles 79$^\circ$ and 139$^\circ$ with respect to the beam direction. The target was made up of 1 mg/cm$^2$ of enriched (96$\%$) $^{100}$Mo backed with a 9 mg/cm$^2$ natural Pb. The $8\times10^8$ two-fold coincidence data were sorted in a symmetric and an asymmetric angle dependent matrix using 
the sorting program INGASORT \cite{bhow1}. The symmetric matrix was analyzed with RADWARE program ESCL8R \cite{rad1,rad2} to build the level scheme of $^{108}$Cd. The present data confirmed the placement of gamma-transitions to the Band 5 of $^{108}$Cd as reported in the reference \cite{kelsall}.

\indent The asymmetric matrix was analyzed using the program DAMM \cite{mil1} to extract the lineshapes of the
 gamma transitions of the Shears Band. The lineshapes 
were observed above the I$^{\pi}$=13$^-$ level and were extracted using the 121 keV gate. The lineshapes of 316 (I$^{\pi}$=14$^-$) and 522 (I$^{\pi}$=15$^-$) keV transitions were also extracted from 677 (I$^{\pi}$=16$^-$) keV gate.
The level lifetimes were estimated by using the LINESHAPE analysis 
code of Wells and Johnson \cite{john1}. The analysis procedure has been described in detail in reference \cite{datta1}.

\begin{figure*}
\includegraphics[scale=0.5]{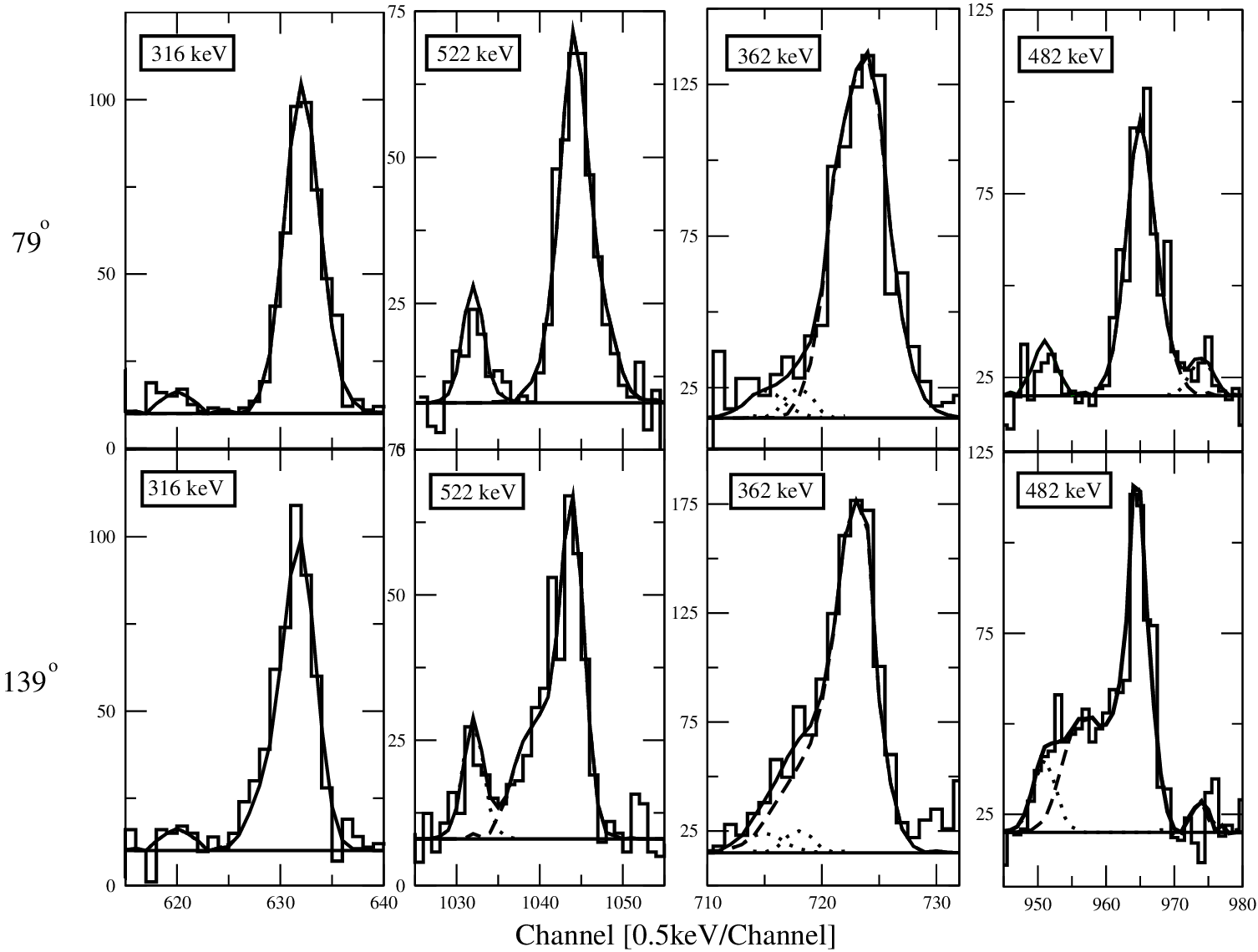}
\caption{\label{fig:fig1}~Experimental and theoretical lineshapes for the 316, 522, 362 and 482 keV $\gamma-$rays of $^{108}$Cd at 79$^\circ$ and 139$^\circ$ with respect to the beam direction. The contamination peaks are shown by dotted lines and theoretical lineshapes are shown by solid lines.}
\end{figure*}

\indent The intensity of the topmost transition namely, 798 keV, was too weak for the lineshape analysis. Thus, the effective lifetime for 21$^-$ was found by fitting the observed lineshape of 704 keV transition by assuming 100$\%$ side-fit. The estimated effective lifetime was 0.80(7) ps. For 20$^-$ level, the effective lifetime of 
21$^-$ level and side-feeding lifetime were considered as input parameters. The side feeding intensity was fixed to reproduce the observed intensity pattern of the band. In this way, 
each lower level was added one by one and fitted until all the seven levels were included in a global
 fit where only the in-band and side feeding lifetimes were allowed to vary. This procedure of global 
fit was repeated for the forward and backward spectra. The lineshapes of 316 and 522 keV transitions in 677 keV gate were fitted by assuming 100$\%$ top-feed with a feeding lifetime equal to the effective lifetime of the $16^-$ level. The uncertainties in the level lifetimes were derived from the behavior of $\chi^2$ fit in the
 vicinity of the minimum. Fig.~\ref{fig:fig1} shows the experimental and fitted lineshapes for four gamma transitions
in this Band namely 316, 522, 362 and 482 keV. The results of the global fit are summarized in the Table~\ref{tab:table1}, where the lifetimes of 14$^-$ and 15$^-$ level were obtained from both 121 and 677 keV gates. It should be noted
that the quoted errors do not include systematic error in the stopping power values which may be as large
as $\pm$20$\%$ \cite{clark1}. The lifetime of the 14$^-$ level has not been reported by Kelsall {\it et al.} \cite{kelsall}. In addition, in the present work the lifetime of the 15$^-$ level has been found to be 0.39(4) ps which differs substantially from the previously reported value of 0.69(6) ps \cite{kelsall}. The lifetimes deduced for all other levels are in good agreement with the previously reported values.

\indent The estimated B(M1) transition rates have been deduced from the standard 
formula \cite{voi} and listed in Table~\ref{tab:table1}. It is evident from the evaluated B(M1) rates that the values decrease as a function of angular momentum till I$^{\pi}$ = 16$^-$ and beyond I$^{\pi}$=17$^-$ there is a distinct increase, followed again by a fall. This observation is very similar to that in $^{196}$Pb \cite{aks} and thus, can be associated with a band crossing in this Shears band.

\indent In order to establish this assumption, the energies and transition rates of the levels of this Shears band have been calculated using the semi-classical model of Shears mechanism \cite{machia1} and compared with observed values. In this model, the Shears angle ($\theta$) is the important variable which can be derived using the equation 
\begin{equation}
cos\theta=\frac{{I_{sh}}^2-{{j}_\parallel}^2-{j_\perp}^2}{2\ {j_\parallel}\ {j_\perp}}
\label{eqn:eqn1}
\end{equation} 
where, $I_{sh}$ is the shears angular momentum. Thus, the Shears angle associated with a specific level depends on the configuration of the band. Kelsall {\it et al.}, have established $\pi[g_{9/2}^{-3}\ g{7/2}]\otimes \nu[h_{11/2}(g_{7/2}/d_{5/2})]$ as the configuration for this band where two of the proton holes are assumed to be antiparallel and therefore do not contribute to the Shears mechanism \cite{kelsall}. 
Thus, in the present work, $\bm{j}_\parallel$ has been assumed to be 3.5$\hbar$ which corresponds to the $g_{9/2}$ proton hole and $\bm{j}_\perp=11.5\hbar$ is determined to reproduce the band head spin of 12$\hbar$. This contribution comes from $\pi g_{7/2}$ and $\nu[h_{11/2}(g_{7/2}/d_{5/2})]$. 
After the band crossing, $\pi[{g_{9/2}}^{-3}g_{7/2}]\otimes$$\nu[{h_{11/2}}^3(g_{7/2}/d_{5/2})]$ configuration has been assigned to the new band 
\cite{kelsall} whose band head spin is 17$\hbar$. In order to reproduce this spin, $\bm{j}_\perp$ has been taken to be 16.5$\hbar$. It is interesting to note that this assumption is in agreement with the experimental alignment gain of $\sim5\hbar$ \cite{thor1}.

\indent Under these assumptions, the shears angle has been calculated using Eq.~\ref{eqn:eqn1} for every angular momentum state of the Shears band and the 
values are listed in Table~\ref{tab:table2}. It is to be noted that the maximum angular momentum for this shears configuration is 20$\hbar$ while
 the band has been observed up to 22$\hbar$. This small difference (10$\%$) due to the core rotation is assumed
 to be a linear function of angular momentum. Thus, $I_{sh}=I-(\Delta R/\Delta I)(I-I_b)$ \cite{clark1}, where $I_b$ is the band head spin 
(=12$\hbar$) and  $(\Delta R/\Delta I)$=0.4 for the present case.

\indent Since the core contribution is small, the energy levels of the band can be calculated following the prescription of Macchiavelli {\it et al.} \cite{machia1}.
\begin{equation}
E_I-E_b=(3/2)V_2\ cos{\theta}_I
\label{eqn:eqn2}
\end{equation} 
where, $E_I$ is the energy of the level with angular momentum I, and ${\theta}_I$ is the corresponding Shears angle and $V_2$ is the strength of interaction between the blades of the shear.

\begin{figure}
\includegraphics*[scale=0.6]{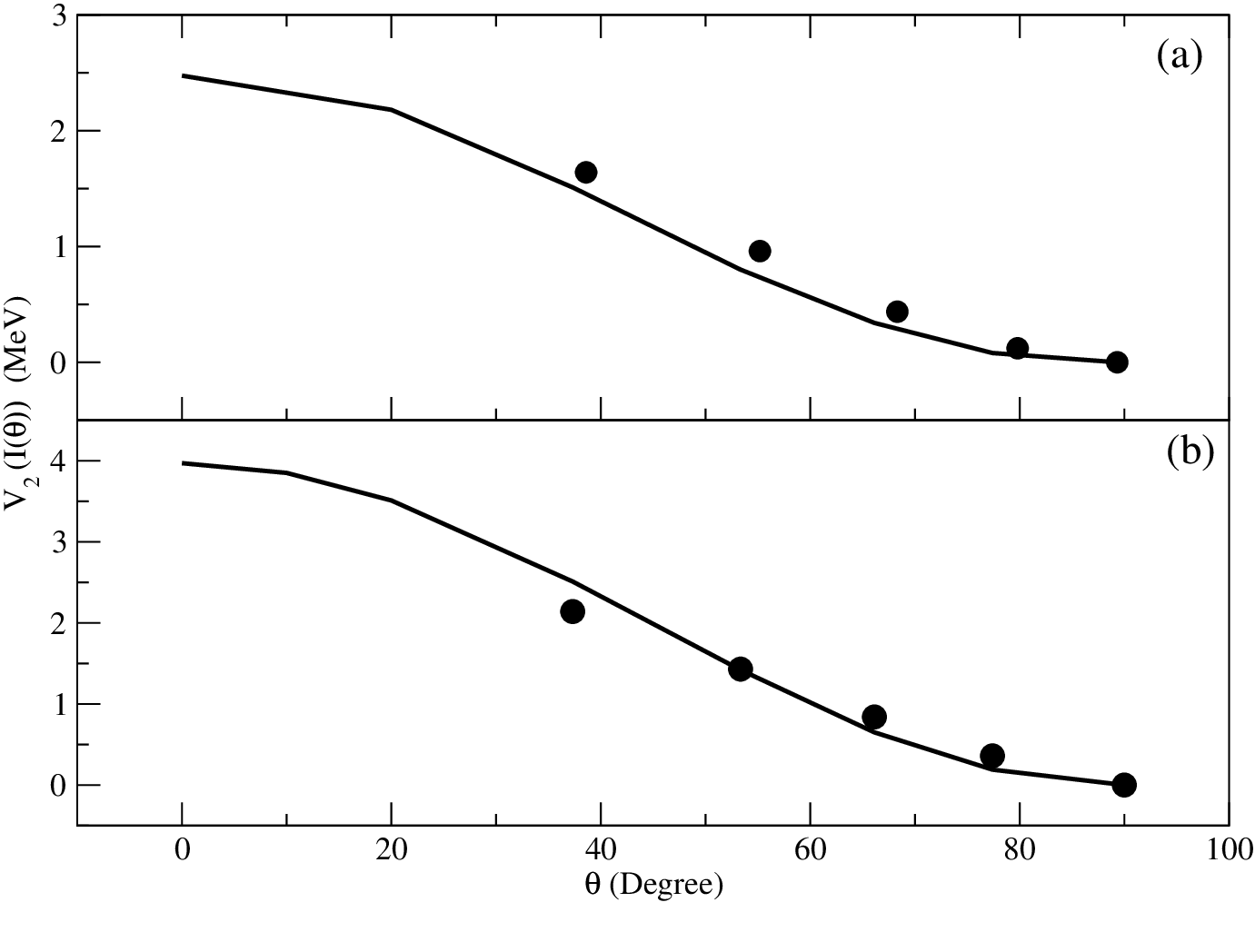}
\caption{\label{fig:fig2}~The effective interaction $V_2$ between the blades of the shear, $\bm{j}_{\parallel}$ and $\bm{j}_{\perp}$, as a function of the shears angle before the band crossing (a) and after (b). The solid line is the fit to the experimental data for (a) $V_2=1.65$ MeV and (b) $V_2=2.65$ MeV.}
\end{figure}

\indent Figure~\ref{fig:fig2}(a) and (b) show the comparison with the data before and after the band crossing, respectively. The experimental level energies are best reproduced for $V_2=1.65$ MeV before the band crossing and $V_2=2.65$ MeV after. It is interesting to note that for the above mentioned configurations, there are three particle-hole combinations before and five combinations after the band crossing. Thus, the interaction strength per pair is $\sim550$ keV. This is in good agreement with the observed systematics of $^{110}$Cd, where $V_2=4$ MeV for eight particle-hole combinations \cite{clark1}. Thus, the observed systematics of $V_2$ support the model of re-opening of the shear angle at the band crossing due to participation of two more particles. This picture may be further clarified by calculating the B(M1) rates.

\begin{figure}
\includegraphics*[scale=0.7]{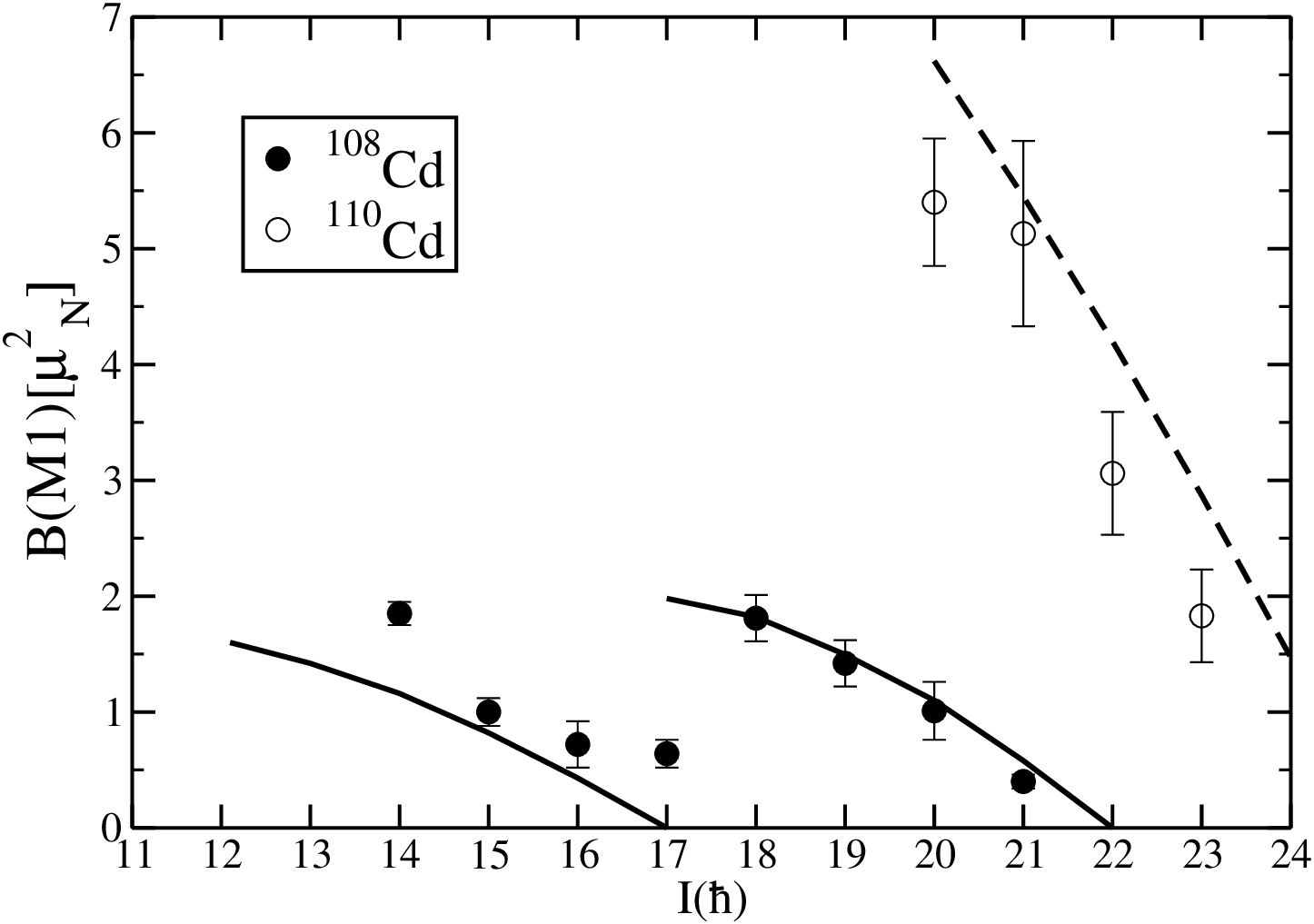}
\caption{\label{fig:fig3}~Experimentally evaluated B(M1) rates as a function of angular momentum in $^{108}$Cd and $^{110}$Cd \cite{clark1}. The value for I$^{\pi}$= 22$^-$ level in $^{108}$Cd has been taken from the reference \cite{kelsall}. The solid and the dashed lines denote the theoretical values from the semiclassical model for $^{108}$Cd and $^{110}$Cd respectively.}
\end{figure}

\indent In the Fig.~\ref{fig:fig3}, the evaluated B(M1) rates have been plotted as a function of angular momentum. These rates can be calculated in the present framework by \cite{clark2}
\begin{equation}
B(M1)=\frac{3}{8\pi} {j_\pi}^2 {g_{eff}}^2 sin^2{\theta}_{\pi}
\end{equation}
where, $g_{eff}=g_{\pi}-g_{\nu}$, $j_{\pi}=3.5$ and ${\theta}_{\pi}$ is related to the shears angle ($\theta$) through 
\begin{equation}
tan\theta_{\pi}=\frac{j_{\nu} sin\theta}{j_{\nu}cos\theta\ +\ j_{\pi}}
\end{equation}
The calculated values of $\theta_{\pi}$ are tabulated in Table~\ref{tab:table2}.
The value of $g_{eff}$ before and after the band crossing was found to be 1.13 and 1.22, respectively. In this calculation, the single particle g-factor of $g_{9/2}$ proton, $g_{7/2}$ neutron and $h_{11/2}$ neutron were taken to be 1.27, -0.21 and 0.21 respectively \cite{datta2} and the normal parity neutron was assumed to have predominantly $g_{7/2}$ character. The calculated values are shown as solid line in Fig.~\ref{fig:fig3}. There is a good agreement between the calculated and observed values and the characteristic variation of the B(M1) rate before and after the band crossing has been well reproduced by the semi-classical model. Thus, the observed B(M1) rates are also consistent with the assumption that the band crossing in the Shears band can be described as the reopening of the shears angle.

\indent It is interesting to note that the previously assigned configuration \cite{thor1} would predict the B(M1) rates to be $\sim$4 times higher as for this configuration $j_\pi=8$ as compared to 3.5 for the present configuration. However, such large B(M1) transition rates have been reported for a Shear band in the immediate even-even neighbor namely, $^{110}$Cd \cite{clark1}. This band was assigned $\pi[{g_{9/2}}^{-2}] \otimes \nu[{h_{11/2}}^2 (g_{7/2}/d_{5/2})^2]$ configuration by comparing the observed B(M1) rates with TAC calculations. In Fig.~\ref{fig:fig3}, the measured values are plotted as open circles. The dotted line represents the calculated values from the present semi-classical model for the above-mentioned configuration and is in good agreement with the observed values. Thus, the B(M1) rates in the Shears bands in this mass region can be used as a marker for the excitation of protons across the N=50 core as they decrease by a factor of four in case of core excitation.

\indent In summery, the observed band crossing in Shears band of $^{108}$Cd can be understood in the following way. The band head spin of 12$\hbar$ is formed by two perpendicular vectors of a shear, $\bm{j}_\parallel=3.5$ and $\bm{j}_\perp=11.5$. The levels up-to I$^{\pi}$=16$^-$, are generated by the closing of the shear angle from 90$^\circ$ to 38$^\circ$. At I$^{\pi}$=17$^-$, this shear closes and a new shear is formed where two more neutrons join $\bm{j}_\perp$. Thus, $\bm{j}_\perp$ increases to 16.5 and the shear angle reopens to 90$^\circ$. The higher spin levels up to 22$\hbar$ are then generated by gradually closing the blades of this new shear. In this picture, it has been assumed that an angular momentum of 2$\hbar$ due to core rotation has been linearly distributed over the levels of this band. Thus, the present work establishes the phenomenon of band crossing in a Shears band in mass-100 region.

The authors would like to thank Professor A. O. Macchiavelli for useful discussion and suggestions. We would also like to thank Professor John Wells for providing the lineshape analysis package and Professor Nimal Singh of Punjab University for providing the target. The authors would also like to acknowledge the efforts of all the technical staffs of the Pelletron facility at IUAC, New Delhi, for smooth operation of the machine.

\newpage

\newpage

\newpage

\newpage
\begin{table}
\caption{\label{tab:table1}~Measured level lifetimes and the corresponding B(M1) transition rates in $^{108}$Cd. The error bars on the measured life-times include the fitting errors and errors in side-feeding intensities.}
\begin{ruledtabular}
\begin{tabular}{cccc}
 $I^{\pi}$ & $E_I$ &  $\tau$ &  B(M1)  \\

             &(keV)&(ps)&(${\mu_N}^2$)\\
\hline
&&&\\
14$^-$ & 316 & 0.95(4) & 1.85(10) \\
15$^-$ & 522 & 0.39(4) & 1.0(12) \\
16$^-$ & 677 & 0.25 (7)& 0.72(15) \\
17$^-$ & 466 & 0.47(6) &0.64(12)\\
18$^-$ & 362 & 0.65 (6)& 1.81(10)\\
19$^-$ & 482 & 0.30 (6)& 1.42(10)\\
20$^-$ & 590 & 0.23(7) & 1.01(15)\\
21$^-$ & 706 & 0.80(7)$^{a}$ & -\\
22$^-$ & 798 & - & -\\
\footnotetext[1]{Effective level lifetime.}
\end{tabular}
\end{ruledtabular}
\end{table} 

\newpage
\begin{table}
\caption{\label{tab:table2}~Calculated shears angle ($\theta$) and proton angle ($\theta_{\pi}$) for states in Band 5 of $^{108}$Cd \cite{thor1}. $j_\perp$=11.5$\hbar$ and $j_\parallel$=3.5$\hbar$ were assumed for $12^- \leq \ I^{\pi} \leq \ 16^-$ and  $j_\perp$=16.5$\hbar$ for higher levels.}
\begin{ruledtabular}
\begin{tabular}{cccc}
$I^{\pi}$& $E_I$(keV)& ${\theta}^{\circ}$ & ${\theta_{\pi}}^{\circ}$  \\
\hline
&&&\\
12$^-$ &- & 89 & 72 \\
13$^-$ &121&80 & 64 \\
14$^-$ &316 & 68 & 54 \\
15$^-$ &522 & 55 & 43 \\
16$^-$ &677 & 38 & 30 \\
\hline
17$^-$ &466 & 88 &76\\
18$^-$ &362 & 77 &66\\
19$^-$ &482 & 66 &56\\
20$^-$ &590 & 53 &45\\
21$^-$ &706 & 37 &31\\
22$^-$ &798 & 0 & 0\\

\end{tabular}
\end{ruledtabular}
\end{table}

\end{document}